\documentstyle[psfig]{mn}

\title{3D numerical simulation of gaseous flows structure
in semidetached binaries}
\author[D. V. Bisikalo et al.]
       {D. V. Bisikalo,
        $^1$\thanks{E-mail address: bisikalo@inansan.rssi.ru}
        A. A. Boyarchuk,$^1$ V. M. Chechetkin,$^2$
\newauthor
        O. A. Kuznetsov$^2$
        \thanks{E-mail address: kuznecov@spp.keldysh.ru}
        and D. Molteni$^3$
        \thanks{E-mail address: molteni@gifco.fisica.unipa.it}\\
       $^1$ Institute of Astronomy of Russian Academy of Sciences,
       48 Pyatnitskaya str., Moscow 109017, Russia\\
       $^2$ Keldysh Institute of Applied Mathematics, 4 Miusskaya
       sq., Moscow 125047, Russia\\
       $^3$ Dipartimento di Scienze Fisiche ed Astronomiche,
       Universit\`a di Palermo,
       36 Via Archirafi, I-90123 Palermo, Italy}
\date{}
\pagerange{\pageref{firstpage}--\pageref{lastpage}}
\pubyear{1998}

\newcommand{\grad}{\mathop{\rm grad}\nolimits}
\begin{document}
\maketitle

\label{firstpage}

\begin{abstract}
    The results of numerical simulation of mass transfer in
semidetached non-magnetic binaries are presented. We investigate
the morphology of gaseous flows on the base of three-dimensional
hydrodynamic calculations in interacting binaries of different
types (cataclysmic variables and low-mass X-ray binaries).

    We find that taking into account of a circumbinary envelope
leads to significant changes in the stream-disc morphology. In
particular, the obtained steady-state self-consistent solutions
show an absence of impact between gas stream from the
inner Lagrangian point $L_1$ and forming accretion disc. The
stream deviates under the action of gas of circumbinary
envelope, and does not cause the shock perturbation of the disc
boundary (traditional `hotspot'). At the same time, the gas of
circumbinary envelope interacts with the stream and causes the
formation of an extended shock wave, located on the stream edge.
We discuss the implication of this model without `hotspot' (but
with a shock wave located outside the disc) for interpretation
of observations. The comparison of synthetic light curves with
observations proves the validity of the discussed hydrodynamic
model without `hotspot'.

    We also consider the influence of a circumbinary envelope on
the mass transfer rate in semidetached binaries. The obtained
features of flow structure in the vicinity of $L_1$ show that
the gas of circumbinary envelope plays an important role in the
flow dynamics, and that it leads to significant (in order of
magnitude) increasing of the mass transfer rate. The most
important contribution to this increase is due to stripping
of mass-losing star atmosphere by interstellar gas flows.

    The parameters of the formed accretion disc are also given
in the paper.

    We discuss the details of the obtained gaseous flows
structure for different boundary conditions on the surface of
mass-losing star, and show that the main features of this
structure in semidetached binaries are the same for different
cases.

    The comparison of gaseous flows structure obtained in 2D and
3D approaches is presented. We discuss the common features of
the flow structures and the possible reasons of revealed
differences.

\end{abstract}

\begin{keywords}
accretion, accretion discs -- binaries: close --  hydrodynamics
-- methods: numerical -- shock waves
\end{keywords}

\section{Introduction}

    The semidetached binaries belong to the class of interacting
stars, where one component fills its critical surface that
causes the mass transfer between components of the system.  In
general, the form of critical surface may be complex (Kruszewski
1963) and special mathematical models are required to describe
the process of mass transfer in such a system (see review of
these models in Lubow 1993). However, in the standard treatment,
semidetached binaries are considered under the assumption that
orbits of components are circular and their rotation is
synchronous with the orbital movement. In this case the critical
surface can be identified with internal surface (Roche surface)
in the restricted three-body problem and it is assumed, that the
mass transfer between components of the system occurs through
the vicinity of inner Lagrangian point $L_1$, where pressure
gradient is not balanced by gravitational force.

    The hydrodynamics of mass transfer through the inner
Lagrangian point $L_1$ has been investigated by many authors.
The detailed analysis of matter flow in the vicinity of $L_1$
was carried out by Lubow \& Shu (1975). Using a perturbation
method they evaluated main characteristics of the flow. In
another approach, based on the analysis of Bernoulli integral,
the stream parameters were specified as well and the dependence
of the mass transfer rate upon the degree of Roche lobe
overfilling was obtained (Paczy\'nski \& Sienkiewicz 1972;
Savonije 1978).

    For adequate description of the mass transfer process in the
binary system besides determination of stream parameters it is
also necessary to consider the further behavior of flowlines
during movement of matter from $L_1$. It is the process of mass
transfer produces the general flow structure and, accordingly,
determines basic observation evidences, therefore the main
attention was paid to study of this question. For the first time
movement of particles leaving $L_1$ and moving in the
gravitational field of binary system was considered by Warner \&
Peters (1972), Lubow \& Shu (1975) and Flannery (1975). These
results were obtained using a simplified ballistic approach for
analysis of the gas movement without taking into account of
hydrodynamic effects. To study the influence of the circumbinary
envelope on gas movement and, accordingly, for a correct
description of the flow, the solving of full system of
hydrodynamic equations is required. This is possible only in the
framework of rather complex mathematical models.

    The use of numerical methods for investigation of
hydrodynamics of mass transfer in semidetached binaries was
limited by the computer power for a long time so 2D models were
used for the analysis of the flow structure. Despite the
restrictions of 2D approach, it allowed to consider some details
of the flow structure correctly and to obtain a set of
interesting results (see, e.g., Sawada, Matsuda \& Hachisu 1986;
Sawada et al. 1987; Taam, Fu \& Fryxell 1991; Blondin, Richards
\& Malinowski 1995; Murray 1996). Last years the possibility of
numerical hydrodynamic simulation of mass transfer in the
framework of more realistic 3D models (Nagasawa, Matsuda \&
Kuwahara 1991; Hirose, Osaki \& Minishige 1991; Molteni,
Belvedere \& Lanzafame 1991; Sawada \& Matsuda 1992; Lanzafame,
Belvedere \& Molteni 1992, 1994; Belvedere, Lanzafame \& Molteni
1993; Meglicki, Wickramasinghe \& Bicknell 1993; Armitage \&
Livio 1996) appeared. In particular, these authors considered
the formation of accreting disc in semidetached binaries
(Nagasawa et al.  1991; Sawada \& Matsuda 1992) and investigated
the interaction of stream of matter leaving $L_1$ with the disc
(Hirose et al.  1991; Armitage \& Livio 1996).

    Unfortunately, many 3D investigations were carried out
during a rather small time-scales and this fact did not allow to
consider the real flow morphology, accordingly, to evaluate the
influence of forming circumbinary envelope on the flow
structure. Some progress in the investigation of general flow
structure in semidetached binaries was achieved in works Molteni
et al. (1991), Lanzafame et al.  (1992, 1994), and Belvedere et
al. (1993), where 3D numerical simulations were carried out on
the sufficiently large time intervals. A set of interesting
results was obtained in these works, however using of method SPH
(Smoothed Particle Hydrodynamics) did not allow to consider the
influence of the circumbinary envelope on the flow structure as
far as the computational restrictions of SPH method did not
permit to investigate flows with large density gradients, and,
accordingly, the account of the influence of circumbinary
envelope on the mass transfer was not quite correct. For the
first time the morphology of gaseous flows in binaries was
accurately considered by authors in Bisikalo et al.  (1997a,b).

    In this work we present the results of 3D numerical study of
the flow structure in semidetached non-magnetic binaries. TVD
(Total Variation Diminishing) method of solving of hydrodynamic
equations used in this paper has allowed to investigate the
morphology of gaseous flows in the system and to consider the
influence of forming circumbinary envelope, despite the presence
of significant density gradients.  The numerical simulations of
mass transfer in semidetached binaries have been conducted on
large time intervals that allowed to consider the main features
of flow structure in steady-state regime.  The earlier
conclusions on the flow structure for low-mass X-ray binary
(Bisikalo et al.  1997a,b) are generalized in present work for
the wider class of objects.

    The paper is structured as follows. In Section 2, the
properties of used physical, mathematical and numerical models
are described.  Section 3 contains the results of numerical
simulations. In this Section the stream-disc interaction,
comparison of synthetic light curves with observations, flow
structure in the vicinity of $L_1$, influence of accepted
boundary conditions, and comparison of the results obtained in
2D and 3D models are discussed. Our conclusions follow in
Section 4.

\section{The model}

\subsection{Physical model}

    The semidetached binaries such as cataclysmic variables
(CVs), low-mass X-ray binaries (LMXBs) and supersoft X-ray
sources (SSS) show a lot of interesting observation evidences.
The observations of CVs light curves and X-ray light curves of
LMXBs offer bright evidences of a complex flow
structure in these systems and allow to make assumptions on the
structure of gaseous flows. In particular, in some cataclysmic
binaries, the most well studied of which is Z Cha, the complex
picture of eclipse (`double eclipse') is observed (see, e.g.,
Hack \& La Dous 1993; Cherepashchuk et al.  1996). For its
explanation the hypothesis of an `hotspot' in interaction
zone between the stream and the disc outer edge was suggested
(Smak 1970).  For a number of X-ray sources such as X1822-371
(`dipping' sources), a significant decrease of the radiation
flux at orbital phase 0.8 is observed. It is explained by a
bulge in the accretion disc created by impact of the stream with
the disc (White \& Holt 1982; Mason 1989; Armitage \& Livio
1996).

    Doubtless, the presence of such appreciable observational
evidences of the complex flow structure in semidetached binaries
requires a detailed study of the gas flow. In this work the
investigations of the flow structure are carried out for two
types of semidetached binaries: 1) with parameters typical for
low-mass X-ray binaries, where main sequence dwarf fills its
Roche lobe and transfers the matter to neutron star, and 2) with
parameters typical for CVs, where the accreting star is a white
dwarf. It is assumed in our model, that magnetic field is
negligible and does not influence the gas flow in considered
systems.

    For the numerical simulations we use systems with parameters
similar to those for X1822-371 (Armitage \& Livio 1996), and for
Z Cha (Goncharskij, Cherepashchuk \& Yagola 1985). For the first
system it is assumed, that the mass-losing component has the
mass $M_1$ equal to $0.28M_\odot$; the gas temperature on the
surface is $T=10^4$ K; the mass of compact star is $M_2= 1.4
M_\odot$; the orbital period of the system is $P_{\rm
orb}=1^d.78$; and the distance between centres of components is
$A = 7.35 R_\odot$.  For Z Cha the following parameters are
adopted: the mass of mass-losing red dwarf is $M_1 =
0.19M_\odot$; the temperature of gas on the surface is $T=5
\times 10^3$ K; the mass of the compact star -- white dwarf is
$M_2= 0.94 M_\odot$; the orbital period of the system is $P_{\rm
orb}=0^d.074$; and $A = 0.78 R_\odot$. We suppose that
mass-losing stars in both models fill their Roche lobes. The
accretor radius for Z Cha is adopted as $R_2= 0.009 R_\odot$,
that is equal to the radius of a typical white dwarf.  For
X1822-371 where the accretor is a neutron star, the computer
facilities do not permit us to resolve the real star size,
therefore we adopt the accretor radius equal to $R_2= 0.05
R_\odot$. It should be noted that the larger accretor radius
adopted for the LMXB case does not influence the flow structure
at distances larger than $R_2$, therefore the obtained results
are correct for all calculated region except the small zone
around neutron star ($r \leq R_2$).

    For the adequate description of the flow structure it is
necessary to take into account the influence of radiative
processes on hydrodynamics. The accurate consideration of
non-adiabatic processes in the numerical model increases the
time of calculations significantly. Therefore, taking into
account our computing environment and the necessity of carrying
out calculations at large time-scales for getting of the
steady-state solution, we simplify the model by using the
adiabatic hydrodynamics and an ideal gas equation of state $P=
(\gamma-1)\rho\varepsilon$ ($P$ -- pressure, $\rho$ -- density,
$\varepsilon$ -- specific internal energy). To consider the
energy losses the ratio of specific heats is assumed to be close
to unity:  $\gamma=1.01$, that corresponds to the
near-isothermal case (Landau \& Lifshiz 1959). Such a technique
for taking into account the radiative losses is well known and
is used in practice repeatedly (see, e.g., Sawada et al. 1986, 
1987; Spruit et al.  1987; Molteni et al. 
1991; Matsuda et al.  1992; Bisikalo et al.  1995).

\subsection{Mathematical model}

    To describe the gas flow, the system of 3D hydrodynamic
equations in integrated form and an ideal gas equation of state
with the ratio of specific heats $\gamma$ = 1.01 is used.  The
system of equations for the fluid element with volume $V$ and
surface $\Sigma$ has the following form:

\[
\frac{\partial}{\partial t} \int\limits_V \rho~dV
+ \int\limits_\Sigma\rho({\bmath{v}}\cdot{\bmath{n}})~d\Sigma = 0
\]
\[
\frac{\partial}{\partial t} \int\limits_V \rho{\bmath{v}}~dV
+ \int\limits_\Sigma\rho{\bmath{v}}({\bmath{v}}\cdot{\bmath{n}})~d\Sigma
+ \int\limits_\Sigma P~{\bmath{n}}~d\Sigma =
\int\limits_V{\rho}{\bmath{F}}~dV
\]
\[
\frac{\partial}{\partial t} \int\limits_V \rho E~dV
+ \int\limits_\Sigma\rho h({\bmath{v}}\cdot{\bmath{n}})~d\Sigma =
\int\limits_V {\rho} ({\bmath{F}}\cdot{\bmath{v}})~dV\,.
\]
The specific external force ${\bmath{F}}$ includes the Coriolis
force, gravitational forces of two point masses, and
centrifugal force, and has the form:

\[
{\bmath{F}} = -\grad\Phi+2\cdot[{\bmath{v}}\times{\bf \Omega}]\,,
\]
where the Roche potential $\Phi$ can be written in the form:

\[
\Phi({\bmath{r}}) = -\frac{G M_1}{|{\bmath{r}} - {\bmath{r}}_1|}
-\frac{G M_2}{|{\bmath{r}} - {\bmath{r}}_2|}
-\frac{1}{2}\Omega^2{\left({\bmath{r}}
-{\bmath{r}}_{\rm c}\right)}^2\,.
\]
Here ${\bmath{v}}$ -- velocity; $h$ -- specific full enthalpy:
$h=\varepsilon + |{\bmath{v}}|^2/2+ P/\rho$; $E$ -- specific
full energy: $E=\varepsilon + |{\bmath{v}}|^2/2$; ${\bf \Omega}
= (0,0,\Omega)$; $\Omega=2\pi/P_{\rm orb}$; ${\bmath{r}}_1$,
${\bmath{r}}_2$ -- the centres of components of the system; and
${\bmath{r}}_{\rm c}$ -- the centre of mass of the system.

    The boundary conditions on the surface of mass-losing star
are determined under the assumption that this star fills its
Roche lobe. To build the boundary conditions on the Roche lobe
we use a standard procedure of solving of Riemann problem
between two regions -- one on the surface of the star and other
corresponding to the nearest computational cell (see, e.g.,
Sawada et al. 1986; Sawada \& Matsuda, 1992). On the surface of
mass-losing star we choose the value of density $\rho = \rho_0$.
We also guess that everywhere on the surface of this component
the gas velocity is directed along the surface normal, and its
value is fixed to be equal to the local sonic velocity $v=c_0$.
For Z Cha we also consider the case, when the gas velocity is
fixed to be zero for all the surface of mass-losing star.

    It should be noted that the boundary value of density on the
surface of mass-losing star has no influence on the solution,
due to scaling of the system of equations with respect to $\rho$
(with simultaneous scaling of $P$). So an arbitrary value of
$\rho_0$ can be accepted in calculations, however, when
considering the particular system with known mass-loss rate, to
determine the real values it is necessary to increase the
calculated values of density in accordance with the scale,
defined by ratio of real value of density on the surface of
mass-losing star to the model one.

    On the surface of the compact star and the outer numerical
boundary the free outflow conditions are used.

    As initial conditions the low-density ambient matter
($\rho\sim10^{-6}\rho_0$) in rest in the rotational frame is
accepted.  Subsequently, this matter is forced out from the
system by the gas injecting from mass-losing star.

\subsection{Numerical model}

    The main question for numerical simulation of the
hydrodynamic models is the choosing of solving method and
appropriate scheme for the system of equations. Among the large
variety of finite-difference schemes the so-called Godunov-type
schemes (Godunov 1959) are considered to be the most exact ones.

\begin{figure*}
\centerline{\hbox{\psfig{figure=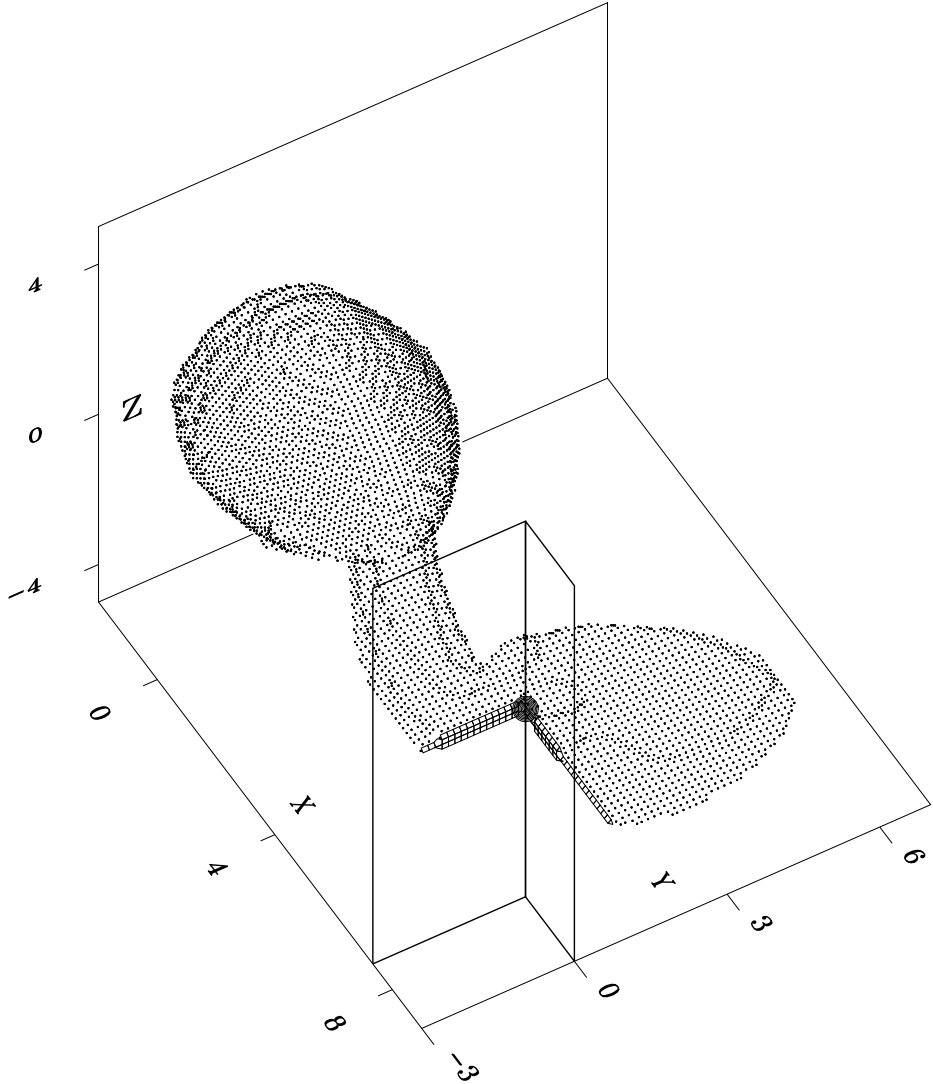,width=6.5in}}}
\caption{3D view of density isosurface at the level
$\rho=0.005\rho_0$. The values of coordinates $X$, $Y$ and $Z$
are in units $R_\odot$. Accretor is marked by the filled circle.
Cross-sections of density isosurface by planes $XZ$ and $YZ$
passing through the accretor are also shown.}
\end{figure*}

    In the present work, we use the modification of explicit TVD
Roe scheme (Roe 1986) for numerical solving of the system of
hydrodynamic equations. The original scheme (first order of
spatial approximation) is modified by monotonic flux limiters in
the Osher's form (Chakravarthy \& Osher 1985) that makes the
scheme of third order of approximation. The special model
simulations show, that the given scheme permits to describe
adequately the flow structure including shock waves and
tangential discontinuities and does not result in artificial
fluctuations and smearing of features of flow. Moreover the used
scheme permits to consider the flows with large density
gradients, that of special importance for consideration of the
influence of circumbinary envelope on the flow
structure.

    The system of hydrodynamic equations is solved in Cartesian
coordinate system, which is predetermined as follows:

  -- the zero of coordinate system is located in the centre of
mass-losing star;

  -- $X$ axis is directed from the centre of mass-losing star to
the accretor;

  -- $Z$ axis is directed along the axis of orbital rotation;

  -- $Y$ is determined to get right-handed coordinate system.

  The computation region is a parallelepipedon \linebreak
$[-A..2A]\times[-A..A]\times[0..A]$ (due to symmetry about the
equatorial plane calculations were conducted only in the top
half-space). Non-uniform difference grids (more fine near the
accretor) containing $78\times 60\times 35$ gridpoints for the
system X1822-371 and $84\times 65\times 33$ gridpoints for the
system Z Cha is used.

\begin{figure*}
\centerline{\hbox{\psfig{figure=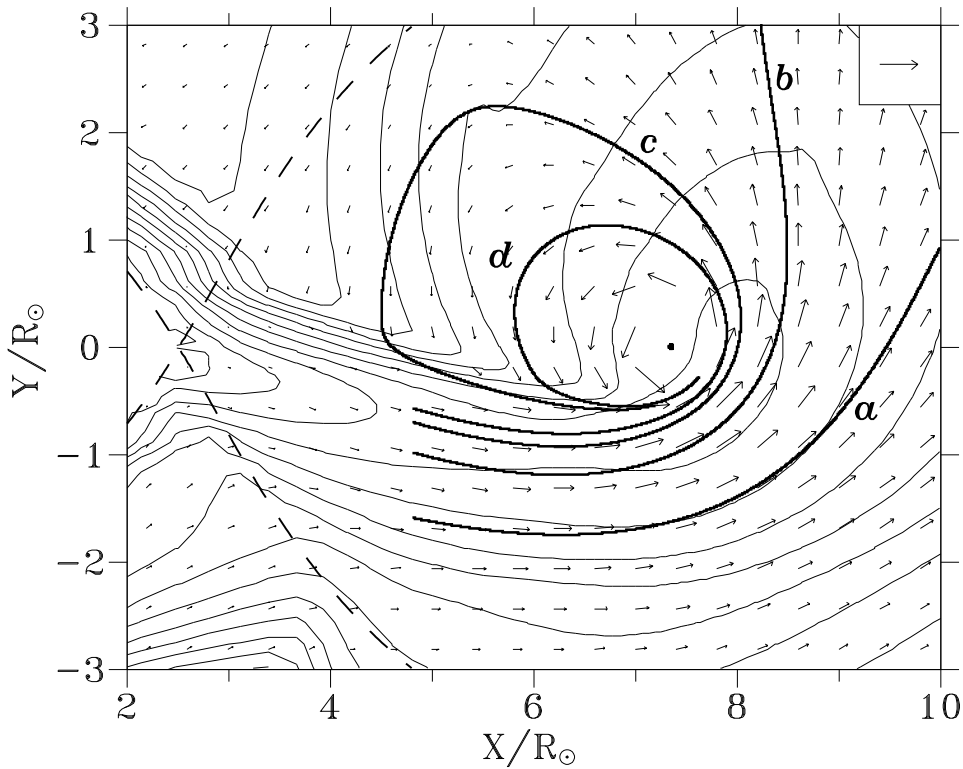,width=5.0in}}}
\caption{Density isolines and velocity vectors in the equatorial
plane of the system.  Roche equipotentials are shown by dashed
lines.  Four flowlines, labelled by markers `$a$', `$b$', `$c$'
and `$d$' are also presented.  The accretor is marked by the
filled circle. Vector in the upper right corner corresponds to
the value of velocity of 800 km s$^{-1}$.}
\end{figure*}

    Solving of the system of equations has been carried out from
initial conditions up to the steady-state regime. To check the
establishment of the steady-state regime we have numerically
monitored flow parameters (density and pressure) as a function
of time. We have checked these parameters inside the spheres
around accretor (with different radii) and inside the sphere
closed to the outer boundary. When the flow patterns do not
depend on the time we suppose that steady-state is reached. To
assure that the obtained solution is steady we have continued
the calculations additionally during 3--5 orbital periods. The
runs have been stopped at 12 orbital period for X1822-371, and
at 20 orbital periods for Z Cha. Characteristic time step in
both runs is approximately $10^{-4}$ orbital period, so the
total number of steps is $\sim 1.2 \times 10^5$ and $\sim 2
\times 10^5$ accordingly. Runs have been conducted on NEC ALPHA
computer (AlphaStation 250 4/266), and CPU time per 1 gridpoint
was approximately equal to $8 \times 10^{-5}$ seconds.  The
total CPU time for both runs is approximately 1 month.

\section{Results and discussion}

    Let us consider the characteristic features of the flow
structure in semidetached binaries, obtained in the framework of
3D hydrodynamic model described in Section 2. As it was
mentioned above, the calculations have been carried out for
typical representatives LMXBs and CVs. The obtained results
testify qualitatively similar nature of the flow in considered
systems, that, in turn, permits to establish the general
character of the steady flow structures for semidetached
non-magnetic binaries.

    Taking into account the qualitative similarity of results,
the general properties of flow structure will be described below
for the X1822-371 system. The results obtained for the system Z
Cha will be used to discuss the quantitative characteristics.

\subsection{Stream-disc interactions}

    The general structure of the gaseous flows, illustrating the
morphology of mass transfer in the system X1822-371, is
presented in Fig.~1, where 3D view of density isosurface at
the level $0.005\rho_{0}$ is shown.  The cross-sections of
density isosurface by planes $XZ$ and $YZ$ passing through the
accretor are also shown. The flow structure presented in Fig.~1
is steady-state and corresponds to a time exceeding 10 orbital
periods. The analysis of presented results allows to reveal the
following features of the flow structure:

  i) the matter of the stream is redistributed into three parts:
the first part forms a quasi-elliptic accretion disc; the second
part moves around the accretor beyond the disc; the third part
of the stream moves towards the external Lagrangian point $L_2$,
then a fraction of this matter leaves the system, while a 
considerable amount of the gas changes the direction of motion 
due to Coriolis force and comes back to the system;

  ii) the interaction between the stream and the disc is
shock-free;

  iii) the stream of matter moving from the vicinity of $L_1$
changes the sizes as it spreads towards the accretor: the
thickness of the stream decreases, and its width in the orbital
plane increases;

  iv) the thickness of the accretion disc is smaller then the
stream thickness.

    A more detailed analysis of the structure of gaseous flows
in the system and evaluation of linear sizes of the disc can be
carried out for the flow structure in the equatorial plane. In
Fig.~2 density isolines and velocity vectors in this plane for
the region with dimensions from 2 to $10R_\odot$ on axis $X$ and
from $-3$ to $3R_\odot$ on axis $Y$ are presented. In Fig.~2
four flowlines are shown as well, labelled by markers `$a$',
`$b$', `$c$' and `$d$'. These flowlines illustrate the
directions of matter flows in the system.

    The analysis of results presented in Fig.~2 verifies the
above conclusion that the part of matter of the stream falls in
the disc at once (flowline `$d$'), and then loses the angular
momentum under the action of numerical viscosity and accretes.
The obtained quantitative evaluations show, that in steady-state
regime the fraction of accreted matter, for the given
semidetached binary, is approximately equal to 75 per cents of
the total amount of matter injected into the system.

    The part of matter, that remains in the system and
influences the flow structure (flowlines `$a$', `$b$', and `$c$'
in Fig.~2) is of special interest. Hereafter we shall name this
part of matter by a circumbinary envelope. It should be noted,
that a significant part of the gas of circumbinary envelope (see
flowlines `$a$' and `$b$') interacts with the matter ejected
from the surface of mass-losing star. The influence of this part
of circumbinary envelope on the flow structure results in
considerable change of the mass transfer regime.  The detailed
analysis of this effect will be presented in subsection 3.3 of
this paper. Another part of circumbinary envelope (see flowline
`$c$') makes a revolution around the accretor and shocks the
stream edge, facing the orbital movement.  This interaction
results in a significant change of the general flow structure in
the system and, in particular, in absence of `hotspot' in
the disc, as well as to the formation of an extended shock wave,
located along the stream edge. Detailed description of the
influence of this part of circumbinary envelope on the
morphology of gas flows in the system is presented below.

    To estimate linear sizes of the disc it is necessary to find
the marginal (`last') flowline along which the matter falls
directly in the disc. Flowline `$d$' in Fig.~2 is the marginal
one and it is easy to determine sizes of the calculated
quasi-elliptic accretion disc: $2.3\times2.0R_\odot$ ($0.31
\times 0.27A$). The thickness of the disc increases with the
distance from the accretor and changes from $\sim 0.05$ to $\sim
0.27R_\odot$ (or from 0.9 to 5.2 accretor radii).  Similar
evaluations of the parameters for quasi-elliptical disc using
the marginal flowline is obtained for Z Cha system as well. For
this system the forming steady-state disc has the size
$0.25\times0.22R_\odot$ ($0.32\times 0.28 A$), and its thickness
varies in the range from $\sim 0.006$ to $\sim 0.04 R_\odot$ (or
from 0.7 to 4.5 accretor radii). It should be noted that for
considered binary systems with approximately equal components
mass ratios, but considerably different various characteristic
parameters (separation $A$ and orbital period),  radii and disc
thickness (adimensionalized using $A$) are approximately
identical.  Moreover, locations of discs in the relation to the
accretor Roche lobe also coincide for different systems.

    The conducted analysis shows, that in all flowlines,
belonging to the disc, up to the marginal flowline `$d$' the
flow is smooth one.  The absence of breaks indicates the
shock-free interaction between the stream and matter of the disc.
There is no a `hotspot' on the disc edge.

    The presented flow structure shows that the stream deflects
under the action of the gas of circumbinary envelope (flowline
`$c$' in Fig.~2), approaches the disc along a tangent line and
does not cause any shock perturbation of the disc edge.

    At the same time the analysis of results shows that the
interaction between stream and circumbinary envelope results in
formation of an extended shock wave, located along the stream
edge turned towards orbital movement. The parameters of this
shock wave, as well as the total energy production in it can be
evaluated from Fig.~3. Here the normalized distribution of the
energy production specific rate $\delta E$ (erg s$^{-1}$
cm$^{-2}$) along the shock wave in the equatorial plane is
presented. In Fig.~3 the boundaries of the shock are shown by
dashed lines: the line on the left shows the starting point of
stream, i.e. the point in the vicinity of $L_1$, the one on the
right -- the ending point of the stream, i.e.  the contact
between the stream and the disc. From the analysis of Fig.~3 we
can see, that the main energy production in the system takes
place in a compact region of the shock wave, located near the
accretion disc. This fact is of great importance for
observational data interpretation as far as the compactness of
the energy production area permits to explain the phase
dependencies of light curve features practically in the same
manner as it was made earlier under the assumption of `hotspot'
existence.

\begin{figure}
\hbox{\psfig{figure=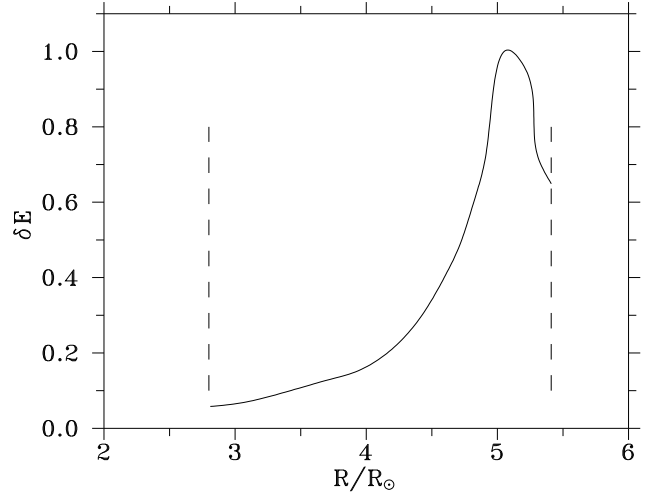,width=3.3in}}
\caption{Normalized to unit distribution of the energy release
specific rate $\delta E$ (erg s$^{-1}$ cm$^{-2}$) along the
shock wave in the equatorial plane of the system. The boundaries
of the shock wave are marked by dashed lines.}
\end{figure}

    The idea of `hotspot' was suggested to explain the observed
light curves of CVs revealing the additional source of
luminosity. It is evident that any new model has to explain this
source of luminosity as well. In the considered model the energy
release occurs in the extended shock wave, located on the stream
boundary. To be convinced of the adequacy of the substitution of
hypothetical `hotspot' by the shock along the stream edge let us
firstly compare an energy release in the hypothetical `hotspot'
and in the shock wave.

    In the model with `hotspot' the energy release due to
interactions between the stream and the disc can be estimated as
(see, e.g., Pringle \& Wade 1985):

\[
\Delta E_{\rm spot} = \frac{1}{4} \frac{G M_2 {\dot
M}}{R_{\rm out}}\qquad (\mbox{erg s}^{-1})\,,
\]
where $\dot M$ is the rate of mass loss by mass-losing star and
$R_{\rm out}$ is the distance between the hypothetical `hotspot'
and the accretor.  In the standard model the mass transfer rate
$\dot M$ is defined as:

\[
\dot M = \rho_0 c_0 S\,,
\]
where $S$ is the stream cross-section in the vicinity of $L_1$
(Lubow \& Shu 1975). Supposing that $R_{\rm out}$ is equal
to the radius of the accretion disc obtained from the
hydrodynamic calculations presented above it is possible to
estimate $\Delta E_{\rm spot}$. The comparison of the energy
release rate in the shock wave $\Delta E_{\rm shock}$ with
$\Delta E_{\rm spot}$ shows, that $\Delta E_{\rm shock}$ exceeds
$\Delta E_{\rm spot}$ by factor 2.  The obtained evaluation of
energy release in cataclysmic binary Z Cha indicates that for
this type of semidetached binaries $\Delta E_{\rm shock}$ is
approximately equal to $\Delta E_{\rm spot}$ and the main part
of energy is also released in a part of the shock close to the
disc. It means that in the considered model without `hotspot'
the value of additional energy release is enough to explain the
observable excess of luminosity. Of course, it is not sufficient
proof of the adequacy of the considered model. The real proof of
the model adequacy can be obtained only from the direct
comparison of the synthetic light curves with observations.

\subsection{Comparison of synthetic light curves with observations}

    At the present time the maximum information on the flow
structure can be obtained from the analysis of light curves of
cataclysmic variables. The well-known humps on the light curves
are observed for cataclysmic variables in quiescent state
(see, e.g., Hack \& La Dous 1993). These humps repeat regular on
the orbital period and have an amplitude up to 1 magnitude (Wood
et al. 1986).  Moreover, for five cataclysmic binaries in
quiescent state (Z Cha, OY Car, V2051 Oph, HT Cas, IP Peg) the
so-called double eclipse is observed.

    To explain these light curves the hypothesis of `hotspot' is
widely used. According to this hypothesis the `hotspot' is
formed as a result of the shock interaction of stream of matter
leaving $L_1$ with the outer edge of the accretion disc (see,
e.g., Smak 1970; Hack \& La Dous 1993; Shore, Livio \& van den
Heuvel 1994). However, as it follows from the numerical studies
of steady-state self-consistent gaseous flows structure the
stream of matter from $L_1$ does not cause the shock
perturbation of the disc. In the hydrodynamic model presented
above the zone of energy release is located outside the disc.
This zone is formed due to the shock interaction between the gas
of circumbinary envelope revolving accretor and the stream of
matter from $L_1$.

    To be convinced of the adequacy of the considered model we
have built the synthetic light curve for cataclysmic variable Z
Cha and have compared it with observations. To obtain the
synthetic light curves the technics described in papers by
Khruzina (1992) and by Khruzina \& Cherepashchuk (1994) is used.
The detailed description of this method adapted for the
considered hydrodynamic model is given in Bisikalo et al.
(1998).

\begin{figure}
\hbox{\psfig{figure=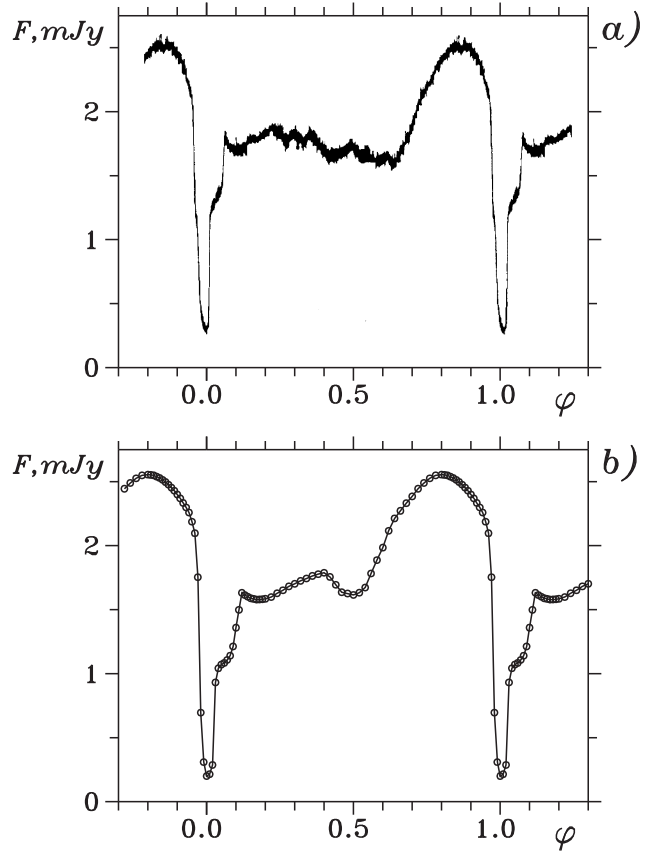,width=3.3in}}
\caption{a) Optical light curve of cataclysmic variable Z Cha
(Wood et al. 1986); b) synthetic light curve of Z Cha obtained
for the hydrodynamic model without `hotspot' (Bisikalo et al.
1998).}
\end{figure}

    Observable and synthetic light curves are presented on
Fig.~4.  The comparison of these curves shows the good
qualitative agreement. Practically all characteristic features
of the observable light curve are repeated on the theoretical
one.  It should be also noted that we have built synthetic light
curves for the different types of cataclysmic variables.  The
comparison of obtained curves (Bisikalo et al. 1998) with
observation shows that in the framework of considered model with
energy release zone located outside the disc it is possible to
explain practically all types of observable light curves.

\begin{figure*}
\centerline{\hbox{\psfig{figure=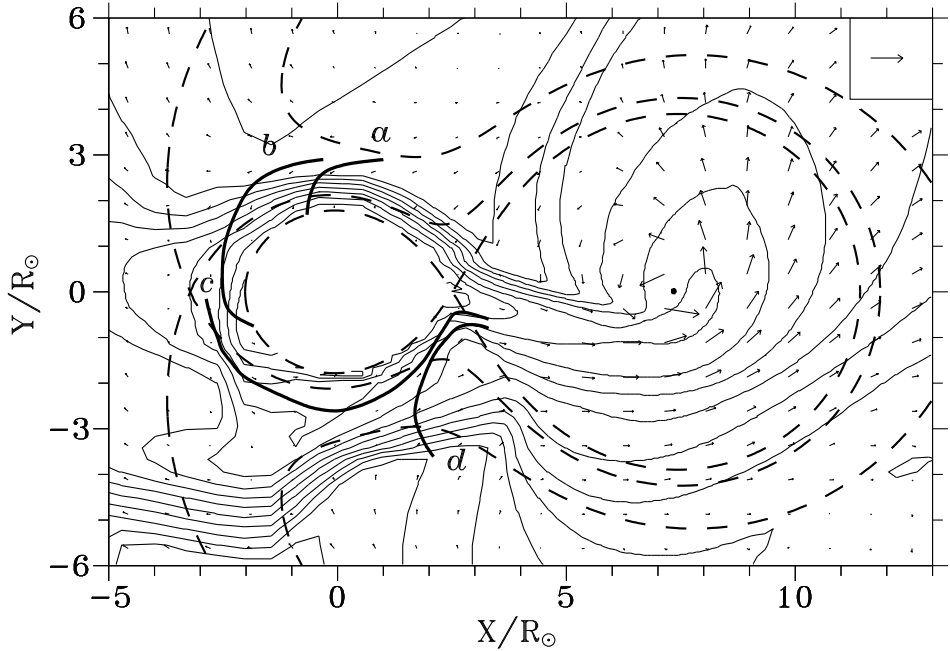,width=5.0in}}}
\caption{The same as Fig.~2 for extended region of the system.
Four flowlines labelled by markers `$a$', `$b$', `$c$' and
`$d$' show the directions of gas movement near the mass-losing
star. Vector in the upper right corner corresponds to the value
of velocity of 800 km s$^{-1}$.}
\end{figure*}

\subsection{Flow structure in the vicinity of $L_1$}

    The stream parameters in the vicinity of $L_1$ were
specified both in analytical (Paczy\'nski \& Sienkiewicz 1972;
Lubow \& Shu 1975; Savonije 1978) and numerical (see, e.g.,
Nazarenko 1993) works. The main characteristics of matter stream
obtained in various works differ only in details, and at
present time, these expressions are widely used as standard for
the mass transfer analysis in semidetached binaries (see, e.g.,
Pringle \& Wade 1985; Shore et al. 1994).  Unfortunately, in all
these works the influence of forming circumbinary envelope
on the flow structure in the vicinity of mass-losing star
was not taken into account. In this work we can consider the
contribution of circumbinary envelope on the basis of 3D
numerical simulations.

    The general flow structure in the equatorial plane of
considered system is presented on Fig.~5, where density isolines
and velocity vectors are presented. This figure is similar to
Fig.~2 but the results are presented for larger region with
sizes from $-5$ to $13R_\odot$ on $X$ axis and from $-6$ to
$6R_\odot$ on $Y$ axis. On Fig.~5 four flowlines (labelled by
markers `$a$', `$b$', `$c$', `$d$'), illustrating the directions
of flows in the system are shown as well.

    The analysis of results presented on Fig.~5 shows, that the
significant part of the gas of circumbinary envelope (flowlines
`$a$' and `$b$') reaches the surface of mass-losing star (Roche
lobe) and prevents the gas to escape from the star surface. A
part of matter of the envelope (flowlines `$c$' and `$d$') blows
away the gas from the surface of mass-losing star and becomes
involved into the process of stream formation.

    The details of the flow structure near the inner Lagrangian
point are shown in Fig.~6, where the same flow parameters as in
Fig.~5 are presented in small region of the equatorial plane
with sizes from 1 to $4R_\odot$ for $X$-direction and from
$-1.5$ to $1.5R_\odot$ for $Y$-direction. Fig.~6 shows that the
gas of circumbinary envelope considerably changes the flow
structure near $L_1$ and, in particular, strips off part of star
atmosphere. In Fig.~6 we can see also the asymmetry of the
influence of circumbinary envelope on the stream. The gas
moving from above -- on the way of orbital movement is accreted
by mass-losing star and only in a small area in the proximity of
$L_1$ it blows out the matter from the surface and transfers it
into the stream.  The gas of the envelope, coming to $L_1$ from
below, strips off the matter from significant part of the
surface.

\begin{figure}
\hbox{\psfig{figure=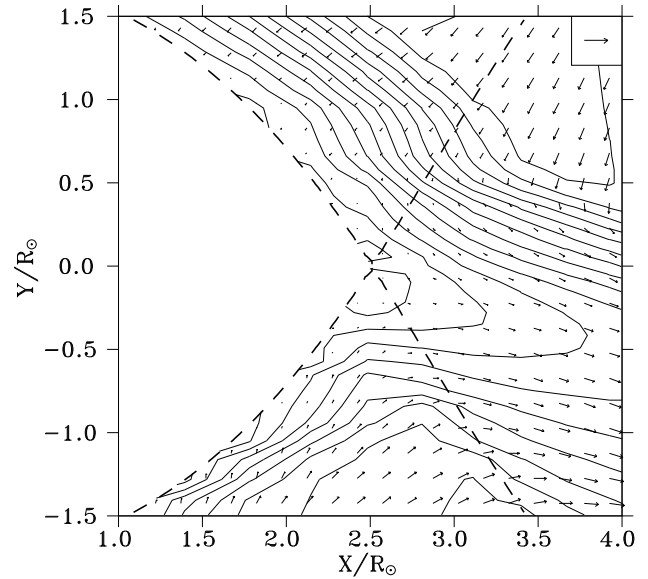,width=3.3in}}
\caption{Density isolines and velocity vectors in the vicinity
of $L_1$ (equatorial plane). Roche equipotential is shown by
dashed lines. Vector in the upper right corner corresponds to
the value of velocity of 300 km s$^{-1}$.}
\end{figure}

    The considered effect of `stripping off' the matter from the
surface of mass-losing star considerably changes the usual point
of view on the mechanism of stream formation and on the mass
transfer parameters. According to the standard model, the
atmosphere structure near the surface of mass-losing star is
determined by the equation of hydrostatic equilibrium (see,
e.g., Lubow \& Shu 1975; Pringle \& Wade 1985). For the adopted
parameters of the system and the temperature (sonic velocity) of
the gas on the mass-losing star surface, the energy of the gas
is not sufficient for the direct escape from the star surface.
Therefore the matter flow connected with thermal escape will be
negligible in comparison with the flow through the vicinity of
$L_1$. The gas of surface layer can flow along the surface
of the star, however in this case the total mass flow to the
system increases slightly (Lubow \& Shu 1975). The situation
changes drastically in the case of taking into account of
a circumbinary envelope, as far as in this case the gas of
surface layer can be `stripped off' and blown away into the
system.

\begin{figure*}
\centerline{\hbox{\psfig{figure=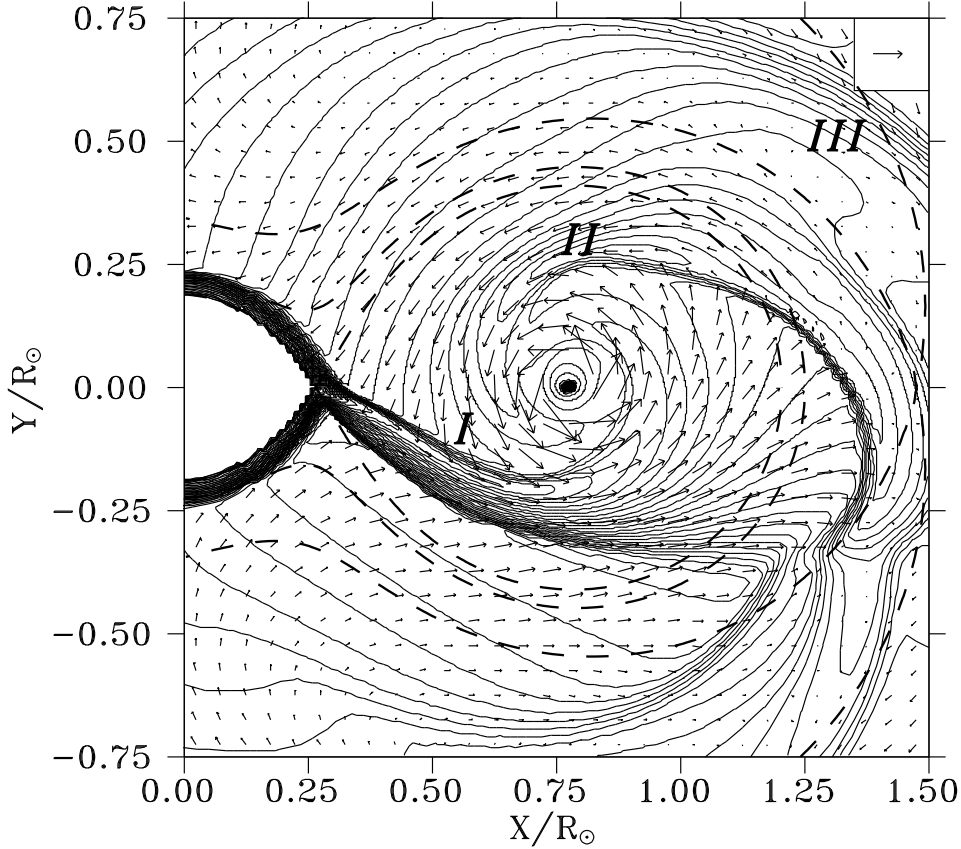,width=5.0in}}}
\caption{a) Density isolines and velocity vectors for Z Cha for
2D run (equatorial plane). Roche equipotentials are shown by
dashed lines. Shock waves are denoted by markers `$I$', `$II$',
`$III$'. Vector in the upper right corresponds to the value of
velocity of 1000 km s$^{-1}$.}
\end{figure*}

\begin{figure*}
\centerline{\hbox{\psfig{figure=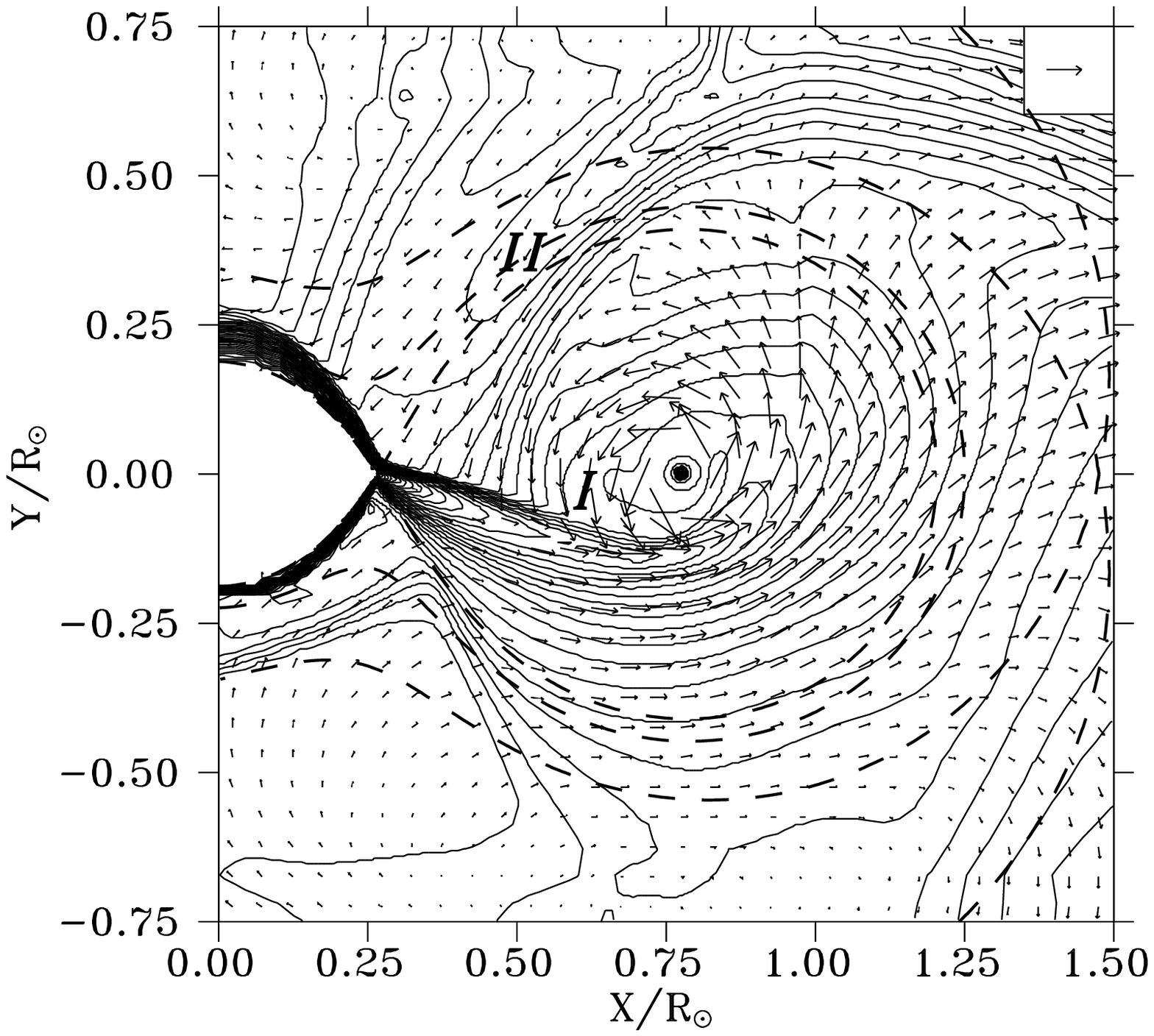,width=5.0in}}}
\contcaption{b) Density isolines and velocity vectors for Z Cha
for 3D run (equatorial plane).  Roche equipotentials are shown
by dashed lines.  Shock waves are denoted by markers `$I$',
`$II$'. Vector in the upper right corner corresponds to the
value of velocity of 1000 km s$^{-1}$.}
\end{figure*}

    On the significant part of the star surface the rarefied gas
of circumbinary envelope has sufficiently large momentum for
capturing the matter from the surface layer. Subsequently this
matter forms the stream and determines the value of the total
mass transfer rate together with the gas leaving the vicinity of
$L_1$. The analysis of results shows, that in the considered
systems X1822-371 and Z Cha the mass transfer rate is one order
of magnitude higher than one predicted by theoretical
estimations, calculated without taking into account the
influence of circumbinary envelope for the same values of gas
parameters on the surface of mass-losing star.

\subsection{Influence of the accepted boundary conditions on the
flow structure}

    The results of calculations for semidetached binaries of
various types show the qualitative similarity of the flow
structure.  Formally, this fact allows to assert, that the
change of parameters of the system does not result in
considerable changes of the flow structure. However, before
making a conclusion about the universal character of the
proposed model, let us consider possible variations of the flow
structure due to changes of boundary parameters of the gas,
injected into the system.

    The study of this problem is stimulated first of all by the
fact that the adopted boundary conditions on the mass-losing
star surface determine the flow structure. Moreover, in view of
limited clarification of the behavior of the surface layers of
star when it fills Roche lobe, there is some
arbitrariness in specifying of the appropriate boundary
conditions.

    As it was mentioned above, to specify the boundary
conditions on the surface of mass-losing star it is necessary to
estimate the temperature, the density and gas velocity. The
first parameter -- temperature is determined from the
observations and its value is known with the satisfactory
accuracy. Second parameter -- density does not influence the
solution due to scaling of the system of equations upon
$\rho$. Therefore, the only boundary condition that
is not well defined and that can influence the flow structure is
the gas velocity.

    The results of calculations for low-mass X-ray binary
X1822-371 and for cataclysmic binary Z Cha have been obtained
under the assumption, that the gas on the mass-losing star
surface has velocity equal to the local sonic velocity, i.e. the
Mach number ${\cal M}$ is equal to 1 and it is the limiting case
of maximum possible value of the boundary gas velocity. Another
limiting case corresponds to zero gas velocity. So, calculations
for Z Cha have been reconducted under the assumption that the
boundary gas velocity is equal to zero, in order to study the
influence of the boundary conditions on the flow structure.

    The results obtained under this assumption show that there
are some insignificant quantitative changes in the flow
structure. In particular: i) the mass transfer rate, in
comparison with the first model (${\cal M}$=1), has decreased
now by 25 per cents; ii) the total energy production in the
shock wave has decreased by 15 per cents. At the same time, the
common flow structure has preserved all characteristic features.
In runs conducted with different boundary parameters we have
obtained the formation of circumbinary envelope which deflects
the stream and leads to the shock-free stream-disc interaction.
This fact allows to conclude that the considered model of the
flow structure for semidetached non-magnetic binaries is an
universal one.

\subsection{Comparison of results obtained in 3D and 2D models}

    For a long time the 2D approach was the main one in
numerical simulations of mass transfer in binaries. Increasing
of computational facilities allows to include the third
dimension into the hydrodynamic models and as a consequence to
make them more realistic.  Unfortunately the 3D models are very
time-consuming and accordingly are not very refined, so the
question of 2D approach applicability is not scholastic one.

    To evaluate the validity of 2D approach we have made the 2D
simulation of the flow structure for Z Cha binary system. To
make the comparison accurately we have used the same parameters
and boundary conditions as in 3D model described above. The
results of calculations for 2D and 3D models are presented on
Figs~7a and 7b accordingly. On these figures density isolines
and velocity vectors in region of the equatorial plane with
sizes from 0.0 to $1.5R_\odot$ for $X$-direction and from
$-0.75$ to $0.75R_\odot$ for $Y$-direction are shown.

    Comparison of Figs~7a and 7b shows that steady-state
flow structures obtained in 2D and 3D models have a set of
common features:

  i) the accretion disc is formed;

  ii) the circumbinary envelope plays an important role in the
formation of the gaseous flows structure;

  iii) the stream-disc interaction is shock-free, and the
`hotspot' does not exist in both models;

  iv) part of the stream revolves around the accretor and
interacts with itself causing the formation of shock wave `$I$'
on the outer edge of the stream;

  v) the interaction of gas of the stream with circumbinary
envelope causes the formation of the shock waves labelled by
marker `$II$'.

    Resuming the above points we may conclude that the
qualitative characteristics of the flow structure in the inner
region for the considered binary are similar. In turn, it means
that 2D model give a rather correct qualitative description of
the gas flow structure in this region.

    As it follows from a comparison of results presented in
Figs~7a and 7b there are also some quantitative differences of
the flow structure in the vicinity of the accretor. In
particular, in 3D case the accretion disc has an elliptic form
and the stream of the matter leaving $L_1$ goes close to the
accretor.  In 2D case the disc is more circular and the angle of
deviation of the stream is greater than in 3D case, therefore
the stream goes far from the accretor.  This fact may be
confirmed by the results presented in Fig.~8, where the density
distributions (normalized on unity) along the line passing
through the accretor parallel to $Y$ axis are shown. From
physical point of view the close moving of the stream in 3D case
is rather evident, because in this case we consider all three
dimensions and gas of the disc can expand on $Z$-direction,
that, in turn, allows the stream to go closer to the accretor.
The distance between the stream and the accretor defines the
ratio between the gas flow leaving the system through the
vicinity of $L_2$, and the gas flow revolving around accretor.
This difference in the gas fluxes leads to the flow structure
changes in the outer regions of the system. In 3D case (where
the stream moves close to accretor) the most part of the stream
is involved into movement around accretor, while in 2D case the
gas outflow through the vicinity of $L_2$ is more effective. In
particular, for 2D case we see that the dominant gas flux
leaving the system through $L_2$ moves around the centre of mass
of the binary system in clock-wise direction and causes the
formation of typical bow shock labelled by `$III$', while in 3D
case this bow shock does not arise.

    It should be noted that the obtained 3D solution was a
steady-state one. The run have been conducted up to 20 orbital
period and we have not found changes in the flow structure. For
2D case the solution is quasi steady-state and even for large
evolution time (few orbital periods) quasi periodic changes of
the flow structure is observed.

    Resuming the comparison between obtained 2D and 3D
solutions, we may say that for the considered case of
$\gamma=1.01$:

  i) the 2D model gives a rather good qualitative description of
the flow structure in the inner regions of the binary, while at
the outer regions the 2D results are not reliable.

  ii) the 2D model does not give an adequate quantitative
description.

\section{Conclusion}

    This work deals with the results of 3D numerical simulation
of the gaseous flows structure in semidetached binaries of
various types. The analysis of results shows the significant
influence of rarefied gas of circumbinary envelope on the
flow patterns in these systems. The gas of circumbinary
envelope interacts with the stream of matter and deflects it.
This leads, in particular, to the shock-free (tangential)
interaction between the stream and the outer edge of forming
accretion disc, and, as the consequence, to the absence of
`hotspot' in the disc.

    At the same time it is shown, that the interaction of the
gas of circumbinary envelope with the stream results in the
formation of an extended shock wave located along the stream
edge. The comparison of synthetic light curves with observations
proves the validity of discussed hydrodynamic model without
`hotspot'.

\begin{figure}
\hbox{\psfig{figure=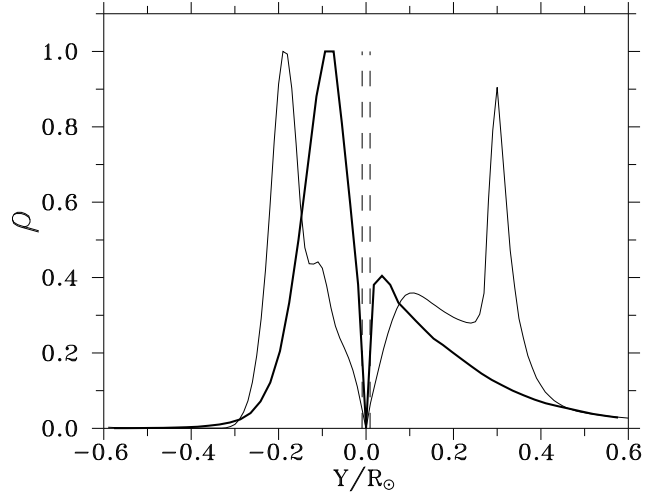,width=3.3in}}
\caption{Normalized density distributions along $Y$ axis passing
through the accretor for 2D (solid line) and 3D (bold line)
cases.  Accretor is placed at the point $Y=0$, and vertical
dashed lines around this point correspond to the accretor size.}
\end{figure}

\begin{figure}
\centerline{\hbox{\psfig{figure=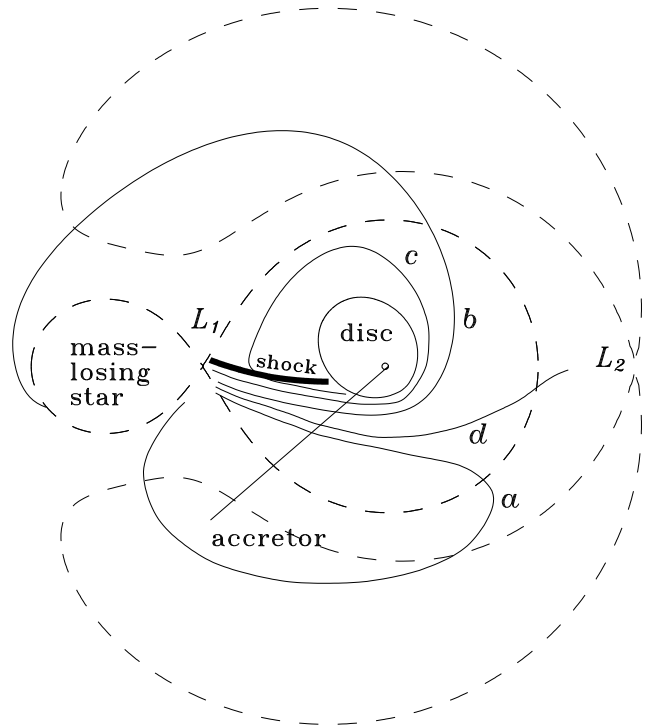,width=3.3in}}}
\caption{Schematic presentation of the main features of flow
structure in semidetached binaries. Roche equipotentials are
shown by dashed lines. Lagrangian points $L_1$ and $L_2$,
location of the accretor, and quasi-elliptic accretion disc are
marked. Shock wave located on the stream boundary is presented
by bold line. Flows labelled by markers `$a$', `$b$', `$c$' and
`$d$' are also shown. Flow `$d$' corresponds to matter leaving
the system through the vicinity of outer Lagrangian point $L_2$.
Flow `$c$' belongs to circumbinary envelope and interacts with
the gas stream resulting in formation of the shock wave. Flows
`$a$' and `$b$' belong to circumbinary envelope as well and
strip off the atmosphere of mass-losing star increasing the mass
transfer rate.}
\end{figure}

    It should be mentioned that taking into account of
circumbinary envelope also leads to a drastic change of the mass
transfer parameters in the system. The calculated mass transfer
rate increases in order of magnitude as compared with values got
from the standard model. Moreover, the gas of circumbinary
envelope changes the flow structure near the surface of
mass-losing component that eventually influences the common
structure of gas flows in the system and consequently affects
the interpretation of observational data.

    The qualitative similarity of the obtained solutions for the
various types of semidetached systems permits to speak about
some universal character of the considered hydrodynamic model
for non-magnetic semidetached binaries. The main features of the
obtained flow structure are summarized on Fig.~9, where gaseous
flows in systems, location of the shock wave, as well as forming
quasi-elliptical accretion disc are presented schematically.

    It should be noted that the presented results are obtained
for the steady-state case. For the non-stationary
(transient) case when morphology of the flow is determined by
external factors and is not self-consistent, more features of
the flow may appear, in particular, arising of the zone of the
disc-stream shock interaction  is possible. For example, if the
disc was formed before the filling by the mass-losing star its
Roche lobe than, after the beginning of mass exchange, the
occurrence of `hotspot' would be possible.  Nevertheless, when
the flow structure will become steady-state we will get the
solution without `hotspot'. Therefore the lifetime of such
formation is of importance. As the characteristic lifetime of
`hotspot' it is naturally to take the interval, during which the
quantity of matter, transferred to the system by the stream,
will become comparable with the accretion disc mass, as far as
after the total replacement of disc matter the solution will be
self-consistent. For the mass transfer rate and accretion disc
parameters typical for semidetached binaries it should be
expected that on time intervals of the order of tens of the
orbital period the solution will become steady-state. It means,
that for the most part of time the gaseous flows structure is
described by the model without `hotspot' as presented above.

   Summarizing, we may conclude, that the correct consideration
of circumbinary envelope in the numerical model of the mass
transfer in semidetached binaries makes possible to find the new
features of the flow structure and thus changes our view on the
morphology of gaseous flows. We should note, that despite
the similarity of observation evidences of the hypothetical
`hotspot' and the shock wave obtained in calculations, the
calculated morphology of matter flow in semidetached binaries
differs from the standard model drastically, that, in turn,
requires the revision of some standard concepts.

\section*{Acknowledgments}

   The present work was supported by Russian Foundation of Basic
Researches (Grant 96-02-16140) and by INTAS (Grant 93-93-EXT).

\bsp

\label{lastpage}

\end{document}